\documentclass[aps,prl,amsmath,amssymb,showpacs,preprintnumbers,nofootinbib,twocolumn]{revtex4}

\usepackage{mathrsfs}
\usepackage{graphicx}
\usepackage[colorlinks=true]{hyperref}
\usepackage{color}

\newcommand{\bw}{\begin{widetext}}
\newcommand{\ew}{\end{widetext}}
\newcommand{\be}{\begin{equation}}
\newcommand{\en}{\end{equation}}
\newcommand{\bea}{\begin{eqnarray}}
\newcommand{\ena}{\end{eqnarray}}

\def\vec{\mathbf}

\def\sslash{s\!\!\!\slash }

\begin{document}

\title{Reply to ``Comment on ``Aharonov-Casher and Scalar Aharonov-Bohm Topological Effects""}

\author{Kai Ma\footnote{Present affiliation: KEK Theory Center and Sokendai, Tsukuba, Ibaraki 305-0801, Japan}}
\email{makainca@gmail.com}
\affiliation{School of Physics Science, Shaanxi University of Technology, Hanzhong 723000, Shaanxi, P. R. China}

\author{Sayipjamal Dulat}
\affiliation{School of Physics Science and Technology, Xinjiang University, Urumqi, 830046, P. R. China}

\begin{abstract}
In this Reply we argue that (i) the Hamiltonian, Eq. (17) in our paper (Phys. Rev. Lett. 108, 070405 (2012)), is definitely Lorentz invariant; (ii) the conditions of generating topological Aharonov-Casher(AC) and Scalar Aharonov-Bohm (SAB) effects are essential and physically meaningful; (iii) the Hamiltonians  both in Phys. Rev. Lett. 74, 2847 (1995) and arXiv:1311.4011 are not suitable to describe the polarized spinor particles.
\end{abstract}
%
%\keywords{Topological phases, Atom interferometer}

\pacs{03.65.Vf, 03.75.Dg, 37.25.+k}

\date{\today}

\maketitle

%%%%%%%%%%%%%%%%%%%%%%%%%%%%%%%%%%%%%%%%%%%%%%%%%%%%%%%%%%%%%%%%%%%%%%%%%%%%%%%%%%%%%%%%%%%%%%%%%%%%%%%%%%%%%%%%%%%%%%%%%%%%%

In a recent paper, Choi and Cho \cite{comment} commented on our paper \cite{uone} and pointed out that (i) our Hamiltonian, Eq. (17), is not relativistic, (ii) then that the conditions we derived are irrelevant for a topological AC and SAB effects, and (iii) conclusively that the non-relativistic Hamiltonian employed by Peshkin and Lipkin \cite{correlation} has the same $U(1)_{mm}$ gauge structure for a fixed spin and then is not wrong, but their incorrect interpretation of the spin autocorrelations led to the incorrect conclusion.

 First, Choi and Cho \cite{comment} obtained their first claim by using the argument that ``Dulat and Ma use the spin projection operator such that the polarization direction of the neutral particle does not vary. However, the spin-state projected Lagrangian $L_+$, Eq. (6), cannot preserve the relativistic invariance, because the spin should undergo a Wigner rotation under a general Lorentz transformation". We can not agree with their arguments. Here we remind that the covariant spin projection operator $\hat\Sigma_\pm(\vec s)=(\mathbb{I} \pm \gamma^5\sslash)/2 $ for an arbitrary spin polarization vector $s^\mu$ satisfies the following relations
$$
\hat\Sigma_\pm(\vec s)\psi_{\pm}(\vec{s}) = \pm 1,~~
\hat\Sigma_\pm(\vec s)\psi_{\mp}(\vec{s}) = 0,
$$
which are invariant under Lorentz transformation \cite{Greiner}. 
When the particle is polarized, it is certainly polarized in any reference frame. This is the basic physical fact and will not change under the Lorentz transformation. 
%So, the polarization vector $s^{\mu}$ always exists. 
One should note that when the wave function changes under Lorentz transformation, the polarization direction specified by  spin vector $s^{\mu}$  changes as well.
%So we should change the projection vector $s^{\mu}$ correspondingly.
The most important thing is that the eigenvalue of the spin projection operator never changes, and we can always separate the Hamiltonian into $H_{+}$ and $H_{-}$, and the mixing between these two parts never occurs. 
In our paper, we provided only the expression of $s^{\mu}$ which undergoes Lorentz boost. 
However, it is implicit that we should change it for the Lorentz rotation.
%And we showed that, the effect due to Lorentz boost is very small.
The definition and description of the covariant spin projection operator is given in many textbooks \cite{Greiner}. In summary, the polarization vector $s^{\mu}$ behaves like a vector under the Lorentz transformation, and our Hamiltonian is relativistic which can be seen from the covariance of our formula, and then the non-relativistically approximated Hamiltonian, Eq. (18), is physical. The derived conditions of the AC and SAB setups are also essential and physically meaningful.

Next, Choi and Cho also argued that the Hamiltonian employed by Peshkin and Lipkin \cite{correlation} is correct because for fixed spin it has the same gauge structure $U(1)_{mm}$. We should firstly point out that when the particle is polarized, the $SU(2)_{spin}$ invariance breaks down to $U(1)_{mm}$, which is one of the most important conclusions in our paper. In our approach, we applied the spin projection at first and then obtained the non-relativistic Hamiltonian. However, the authors made the non-relativistic approximation at first and then the spin projection. Not surprisingly, the gauge structure was the same. Since the non-relativistic approximation does not change the underlying gauge structure, it just projects out the eigen-states of the energy projection operator $\Lambda_{\pm}$.  However, these two approaches have distinctive properties. Firstly, in our approach, we saw clearly that the AC and SAB effect were related by Lorentz transformation. Secondly, by using the non-relativistic approximation we showed that when the spin was polarized, the fluctuation was proportional to $\vec{\sigma}\cdot\vec{\mathcal {B}}$ and $\vec{\sigma}\cdot(\vec{\nabla}\times\vec{\mathcal {E}})$. Just because of this and the polarization must be stable in order to get topological phase, both $\vec{\mathcal {B}}$ and $\vec{\mathcal {E}}$ should be zero. Then the fluctuations of polarization are completely negligible. Factually, it is also true in the perfect AC and SAB setups. Our conclusions are consistent with the experimental observations. It is well know that the Pauli Hamiltonian is correct in the non-relativistic limit. The statement ``Peshkin and Lipkin's Hamiltonian is wrong" just means that their Hamiltonian is not suitable to describe the polarized spinor.

  At last, the authors of the comment argued that Peshkin and Lipkin's conclusion is wrong because of their incorrect interpretation for the spin autocorrelations. We should point out that this is again one of our conclusions in our paper \cite{uone}. And we explained in detail in our paper that the quantum fluctuations are completely negligible. So the comment of these authors does not make any sense on our paper.

In conclusion, we don't think that the letter \cite{comment} does carry any essential comments on our paper \cite{uone}. The Hamiltonian in our paper is definitely Lorentz invariant. The Hamiltonian employed by Peshkin and Lipkin \cite{correlation} is incorrect in the case when the particle is polarized. The quantum fluctuations are certainly negligible, as we have explained in our original paper and in this reply as well. The theoretical arguments presented in Ref.\cite{comment} are not convincing because the analysis almost completely ignored the Lorentz transformation on the polarization vector $s^{\mu}$ and their Hamiltonian is also not suitable to describe the polarized spinor particles.

\section{Acknowledgments}
K. M. is supported by the China Scholarship Council and by the Hanjiang Scholar Project of Shaanxi University of Technology. S. D. is supported by the National Natural Science Foundation of China under the Grant No. 11165014.

\end{document}